\documentclass[%
 reprint,
 amsmath,amssymb,
 aps,
prb,
]{revtex4-1}

\usepackage{graphicx}
\usepackage{dcolumn}
\usepackage{bm}

\newcommand{\ket}[1]{\ensuremath{\left| #1 \right>}}

\newcommand{\Tr}{\text{Tr}}
\usepackage{bbold}
\begin{document}


\title{Simplified Thermodynamics for Quantum Impurity Models}

\author{Colin Rylands} 
\email{rylands@physics.rutgers.edu}
\author{Natan Andrei}
\affiliation{Department of Physics, Rutgers University,
Piscataway, New Jersey 08854.
}

\date{\today}

\begin{abstract}
Quantum impurity models play an important role in many areas of physics from condensed matter to AMO and quantum information. They are important models for many physical systems but also provide key insights to understanding much more complicated scenarios. In this paper we introduce a simplified method to describe the thermodynamic properties of integrable quantum impurity models. We show this method explicitly using the anisotropic Kondo and the interacting resonant level models. We derive a simplified expression for the free energy of both models in terms of a single physically transparent integral equation which is valid at all temperatures and values of the coupling constants. 
\end{abstract}

\maketitle

\section{Introduction}

Quantum impurity models (QIM) are ubiquitous throughout physics describing  many experimental systems, from quantum dots coupled to electronic leads and kinks in carbon nanotubes to isolated atoms in wave guides to name but a few. Impurity  systems consist of a bath of particles, free or interacting, coupled to a small localized system of few degrees of freedom, the  impurity. Their apparent simplicity  often belies the intricate strong correlation physics that is at play.
Examples include the Kondo model \cite{RMP,TWAKM} where the bath is a non interacting Fermi liquid and the impurity is a single isolated spin or the Kane-Fisher model\cite{KF} where the bath consists of a 1D gas of interacting fermions and the impurity is a featureless localized potential which causes backscattering.  Such systems  can also provide insight to the more complicated problems such as heavy fermion systems \cite{hewson} or other correlated systems through the use of dynamical mean field theory \cite{dmft}.

 Many QIMs, in particular the examples cited above,  are in fact exactly solvable though Bethe Ansatz \cite{RMP,TWAKM, ryl} . This method provides one with all the eigenstates and eigenvalues of the Hamiltonian which can then lead to a complete analytic description of the physics of these models. The finite temperature thermodynamics often is then expressed in terms of a large, possibly infinite set of coupled integral equations resulting from summing over the many types of excitation of the model \cite{takahashi}. These Thermodynamic Bethe Ansatz (TBA) equations yield a wealth of information but require detailed numerical analysis to be fully studied. For integrable bulk models these TBA equations have then been reformulated, typically using crossing symmetry, to a form which is very convenient for analytic study\cite{DDV1,KP}.

In this work we extend this approach to  quantum impurity models and 
 present a simplified method to describe their thermodynamics. The method is quite generic and can be applied to all known integrable impurity models. We will show the method explicitly by computing the exact free energy of the anisotropic Kondo model and find that it can be expressed in terms of a single non linear integral equation when there is no external magnetic field.

The paper is organized as follows; in section II we introduce the anisotropic Kondo model, briefly describe how its solution is obtained through Bethe Ansatz and also define the transfer matrix, $\tau(u)$ which is the central object of the method. In section III we show how $\tau(u)$ is related to the partition function of the model and derive a single non linear integral equation which governs the thermodynamics of the system. In section IV we perform some checks on this expression and find the low and high energy behaviour of the impurity. In the next section we comment on the isotropic limit of the free energy as well as the extension to non zero magnetic field. In section VI we discuss the interacting resonant level (IRL) model for which there is no transfer matrix, and show how to relate it to the anisotropic Kondo model and how to derive the simplified thermodynamics in such models where the transfer matrix is absent.

\section{Anisotropic Kondo Model}
The archetypal quantum impurity model is the Kondo model which describes a bath of non interacting fermions coupled to a single magnetic impurity. We consider here  the anisotropic Kondo model (AKM) as an example, however the method is generic and works for multichannel versions as well as other integrable QIMs. The Hamiltonian is,
\begin{eqnarray}\nonumber
H=-i\sum_{a=\uparrow\downarrow}\int \psi^\dag_{a}(x)\partial_x\psi_a(x)+J_\parallel\psi_{a}^\dag(0)\psi_{a}^\dag(0)\sigma^z_{aa}\sigma^z_0\\\label{Hkondo}
+\,J_\perp\left(\psi_{\uparrow}^\dag(0)\psi_{\downarrow}^\dag(0)\sigma^+_0+\psi_{\downarrow}^\dag(0)\psi_{\uparrow}^\dag(0)\sigma^-_0\right).\;\;\;
\end{eqnarray}
Here $\psi^\dag_a(x)$ are right moving fermions with spin $a=\uparrow,\downarrow$   coupled  via  exchange interaction to a  local  magnetic impurity described by the Pauli matrices $\vec{\sigma}_0$. When  the two coupling constants,  $J_\perp$ and $J_\parallel$,  are equal the model is isotropic and $SU(2)$  invariant. The Kondo model  exhibits many remarkable features for such a simple model including spin-charge separation, a dynamically generated energy scale and asymptotic freedom, but it is also an integrable model which allows for a complete description of these properties.   Its solution has been obtained  using coordinate  Bethe Ansatz in the region $0\leq J_\perp\leq J_\parallel$ \cite{RMP,TWAKM}. We now briefly review the solution of the model and refer the reader to \citenum{RMP,TWAKM} for a full account.

The exact $N$ particle wavefunction with energy $E=\sum_j^Nk_j$ is written as an expansion over plane waves of the form 
\begin{eqnarray}\nonumber
\sum_{Q,\vec{a}}\int\mathrm{d}\vec{x} ~\theta(x_Q)\left[A_Q\right]^{a_1\dots a_N;a_0}\prod_j^Ne^{ik_jx_j}\psi_{a_j}^\dag(x_j)\ket{0}\otimes\ket{a_0}
\end{eqnarray}
where $a_j=\uparrow,\downarrow$ is the spin of the $j^{\text{th}}$ particle, $Q$ are orderings of the $N$ particles and the impurity in configuration space, $\theta(x_Q)$ is a Heaviside function which is non zero only for a certain ordering, e.g for the trivial ordering, $Q=\mathbb{1}$ we have $\theta(x_1<x_2\dots <x_N)$ and $\ket{a_0}$ is the impurity state. The amplitudes in each region $\left[A_Q\right]^{a_1\dots a_N;a_0}$ are all related to $\left[A_\mathbb{1}\right]^{a_1\dots a_N;a_0}$ by products of the bare S-matrices of the AKM which themselves are obtained from the solution of the Schr{\"o}dinger equation. The S-matrix between the $i^{\text{th}}$ and $j^{\text{th}}$ particle is the denoted $S^{ij}$ while $S^{i0}$ denotes the S-matrix of the $i^{\text{th}}$ particle going past the impurity. Both types can be expressed in terms of single matrix $S^{ij}=R_{ij}(0)$, $S^{i0}=R_{i0}(c)$ with
\begin{eqnarray}\label{r}
R_{ij}(x)=\begin{pmatrix}
1&0&0&0\\
0&\frac{\sinh{(x)}}{\sinh(x-i\gamma)}&\frac{\sinh{(-i\gamma)}}{\sinh(x-i\gamma)}&0\\
0&\frac{\sinh{(-i\gamma)}}{\sinh(x-i\gamma)}&\frac{\sinh{(x)}}{\sinh(x-i\gamma)}&0\\
0&0&0&1
\end{pmatrix}
\end{eqnarray}
The parameter $\gamma$ encodes the anisotropy of the model and is related to the bare couplings $J_\parallel$ and $J_\perp$. For small anisotropy the relationship is universally given by  $\gamma^2\approx J_\parallel^2-J^2_\perp$ but otherwise depends on the cutoff scheme used. Despite this we can unambiguously say that $\gamma\to 0$ gives the isotropic Kondo model and $\gamma=\pi/2$ gives the Toulouse line. The parameter $c^2\approx J^2_\parallel/\gamma^2$ present in the impurity S-matrix is related to the Kondo scale which is defined as $T_K=De^{\pi c/\gamma}$ where $D=N/L$ with $L$ being the system size. The quantity $D$ is the cutoff of the theory which is necessary due to the linear dispersion of \eqref{Hkondo} and its presence means the wavefunctions are constructed with $k_j>-\pi D$. In the end must take the thermodynamic limit, $N,L\to\infty$ as well as $D\to\infty$ while  holding $T_K$ fixed to obtain universal results. We refer to this second limiting procedure as the universal limit.

With the exact wavefunction in hand the spectrum of the theory is obtained by placing the system on a ring of size $L$ with periodic boundary conditions which results in an eigenvalue problem
\begin{eqnarray}\label{eigen}
e^{-ik_jL}\left[A_\mathbb{1}\right]^{a_1\dots a_N;a_0}=\tau(0)^{a_1\dots a_N;a_0}_{b_1\dots b_N;b_0} \left[A_\mathbb{1}\right]^{b_1\dots b_N;b_0}
\end{eqnarray}
where the operator $\tau(0)$, called the transfer matrix, takes the $j^{\text{th}}$ particle around the ring past all the others. It is defined to be
\begin{eqnarray}\label{tau}
\tau(u)=\Tr_{\bar{0}}\left[R_{0\bar{0}}(u-c)R_{1\bar{0}}(u)\dots R_{N\bar{0}}(u)\right]
\end{eqnarray}
where $\bar{0}$ denotes an auxiliary space over which the trace is performed and $0,1\dots N$ indicate the spaces of the impurity and each of the $N$ particles respectively. The eigenvalues of $\tau(u)$ evaluated at $u=0$  yield the single particle momenta, $k_j$ and the energy, so that the exact solution of the model relies on the ability to diagonalize the transfer matrix. Using the quantum inverse scattering method (see for example ref. \citenum{Sklyrev} and references therein)  this  can be achieved with the result that
\begin{eqnarray}\label{eigen2}
e^{-ik_jL}=\prod_{l=1}^{\mathcal{M}}\frac{\sinh{(\lambda_l+i\gamma/2)}}{\sinh{(\lambda_l-i\gamma/2)}}
\end{eqnarray}
where the $\mathcal{M}$ parameters $\lambda_l,\, l=1...\mathcal{M}$, the Bethe roots,  contain the information about the spin degrees of freedom of the model. These roots  are given as solutions of the Bethe Ansatz equations,
\begin{eqnarray}\nonumber
\left[\frac{\sinh{(\lambda_l-i\gamma/2)}}{\sinh{(\lambda_l+i\gamma/2)}}\right]^N\frac{\sinh{(\lambda_l-c-i\gamma/2)}}{\sinh{(\lambda_l-c+i\gamma/2)}}\\\label{BAE}
=\prod_{j\neq k}^{\mathcal{M}}\frac{\sinh{(\lambda_l-\lambda_k-i\gamma)}}{\sinh{(\lambda_l-\lambda_k+i\gamma)}}.
\end{eqnarray}
with ${\mathcal{M}}\leq(N+1)/2$ being an integer related to the total $z$-component of spin of the system, $S^z=(N+1)/2-\mathcal{M}$. Taking the $\log$ of \eqref{eigen2}, inserting the solutions of \eqref{BAE} and summing over the resulting single particle momenta $E=\sum_j^Nk_j$ we obtain the spectrum of the cutoff AKM for finite $N,L$ and $D$. Thereafter, upon taking the thermodynamic and universal limits, see below, we obtain the exact spectrum of \eqref{Hkondo}.  One then obtains the partition function of the model by summing over the spectrum $E$,  $\mathcal{Z}=\sum_{E} e^{-\beta E}$, and proceeding via the Yang-Yang approach to obtain the TBA equations\cite{takahashi}. In the next section we  develop a simpler  approach generalizing to impurity models the approach of Destri and De Vega\cite{DDV2} who considered bulk models with Lorentz symmetry.

\section{Thermodynamics from the transfer matrix}
The importance of the transfer matrix, $\tau(u)$, in the solution of the Kondo problem was highlighted in the previous section. Its eigenvalues evaluated at $u=0$ give the Bethe  momenta $k_j$  at finite $N$ and $L$. In this section we exploit this equivalence between the transfer matrix and the AKM Hamiltonian to derive a simplified and more general expression for the thermodynamics of the model. As mentioned before the method is quite general and can be applied to all other known integrable QIMs.

\begin{figure}
    \includegraphics[trim={20mm 20mm 15mm 5mm},clip]{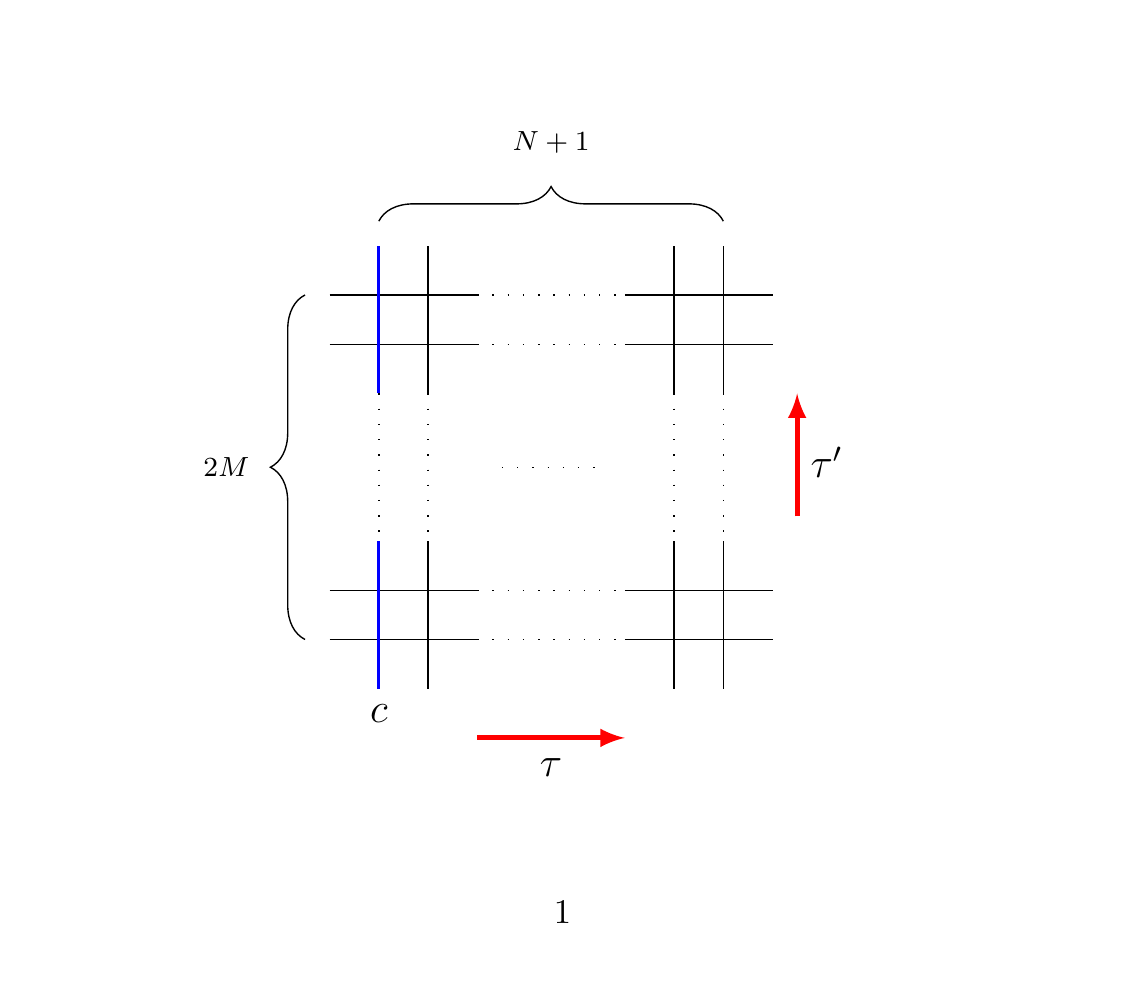}
    \caption{The transfer matrix $\tau(u)$ defined in  \eqref{tau} is also the horizontal transfer matrix of the classical six vertex model. The partition function of \eqref{part1} is that of a model on a square $(N+1)\times 2N'$ lattice with a inhomogeneity of $c$ on the first vertical link and periodic boundary conditions in both directions. Alternately one can calculate the partition function using the vertical transfer matrix $\tau'$ instead. }
    \label{fig:1}
\end{figure}

Returning to the eigenvalue problem \eqref{eigen} we note that it is the same for all particles, with single particle momenta  given by \eqref{eigen2},
\begin{eqnarray}
k_j&=&\frac{2\pi }{L}n_j+\frac{1}{L}\sum_l^{\mathcal{M}}\phi(\lambda_l,\gamma/2)\\
\phi(x,z)&=&i\log{\frac{\sinh{(x+iz)}}{\sinh{(x-iz)}}}
\end{eqnarray}
where $n_j$ are integers which must be distinct for all $j$ due to Fermi statistics. As each  energy eigenvalues is given as the sum over all single particle momenta, $E=\sum_jk_j$, we see that  it splits into two parts, $E=E_c+E_s$. The first, $E_c=\sum_j2\pi n_j/L$,  is due to the charge degrees of freedom and merely describes a free spinless Fermi gas. The second, $E_s=N/L\sum_l^{\mathcal{M}}\phi(\lambda_l,\gamma/2)$,  describes the spin degrees of freedom and is non trivial. Summing in the exponential over all single particle momenta in \eqref{eigen} reduces to the contribution to the energy from the spin degrees of freedom,  $e^{-iEL}=e^{-i\sum_j k_j L}= e^{-iE_sL}$ . This amounts to taking a power of $\tau(0)$
\begin{eqnarray}
e^{-iEL}A_{\mathbb{1}}=e^{-iE_sL}A_{\mathbb{1}}=\tau(0)^NA_{\mathbb{1}}
\end{eqnarray}
as the trivial charge part  cancels out since  $e^{-iE_cL}=1$.

Proceeding along these lines  we can introduce the time variable $t$ and  write $\Tr\{e^{-iHt}\}=\Tr\{e^{-iH_st}\}\Tr\{e^{-iH_ct}\}$, where $H_{s,c}$ refer to the effective spin and charge Hamiltonians \footnote{This can be explicitly done using bosonization}. Further, as  $Tr\{e^{-iH_s t}\}$ is determined only by the eigenvalues of $H_s$, which are in turn  determined by the transfer matrix, we have  
\begin{eqnarray}
\Tr\{e^{-iH_st}\}=\lim_{\text{Univ}}\lim_{N\to\infty}\Tr\{\tau(0)^{2M}\}
\end{eqnarray}
with  $t = \frac{2M}{N} L $ where we first take the thermodynamic limit, $N\to\infty$ holding $D$ fixed and then take the universal limit $D\to\infty$, $M\to\infty$ such that both $t$, and $T_K$ are held fixed. We thus  see  that the transfer matrix provides a regularization of the time evolution operator for the spin part of the AKM at finite $N$ and $L$. 

Carrying out a Wick rotation to imaginary time we  obtain the partition function of the AKM,
\begin{eqnarray}\label{part1}
\Tr\{e^{-\beta H}\}=\mathcal{Z}_c\lim_{\text{Univ}}\lim_{N\to\infty}\Tr\{\tau(0)^{2M}\}\left.\right|_{t\to-i\beta}
\end{eqnarray}
with $\mathcal{Z}_c$  being the charge part of the partition function which is easily computed through standard techniques, see eqn\eqref{fbulk}. In what follows we are concerned only with the impurity properties which carry no charge degrees of freedom and so we will drop  $\mathcal{Z}_c$ from now on with the understanding that we are considering only the spin part of the model.

The quantity $\Tr\{\tau(0)^{2M}\}$ is actually the partition function of the classical 6 vertex model on a $(N+1)\times 2M$ square lattice  wherein $\tau(0)$ is the transfer matrix in the horizontal direction \cite{Baxter}, see FIG. \ref{fig:1}. Due to rotational invariance  we can also compute the partition function of this model using the transfer matrix in the vertical direction which is of a similar form,
\begin{eqnarray}\label{part2}
\Tr\{\tau(0)^{2M}\}=\Tr\{\tau'(0)^N \tau'(-c)\}\\
\tau'(u)=\Tr_{\bar{0}}\left[R_{1\bar{0}}(u)\dots R_{2M\bar{0}}(u)\right].
\end{eqnarray}
The two transfer matrices $\tau'(0)$ and $\tau'(-c)$ commute and are therefore simultaneously diagonalizable. Furthermore, as the limit $N\to\infty$ of the partition function is taken first, $\Tr\{\tau'(0)^N \tau'(-c)\}$ is determined solely by the largest eigenvalue of $\tau'(u)$\cite{Baxter} see FIG. 2. 
The task of computing the partition function of the AKM has therefore been reduced to finding the largest eigenvalue of a six vertex model on a torus - a decidedly simpler task. The only complication being that we must compute this eigenvalue for finite $M$ and in the end take the appropriate limits. 
\begin{figure}
    \includegraphics[trim={20mm 10mm 15mm 5mm},clip,width=.5\textwidth]{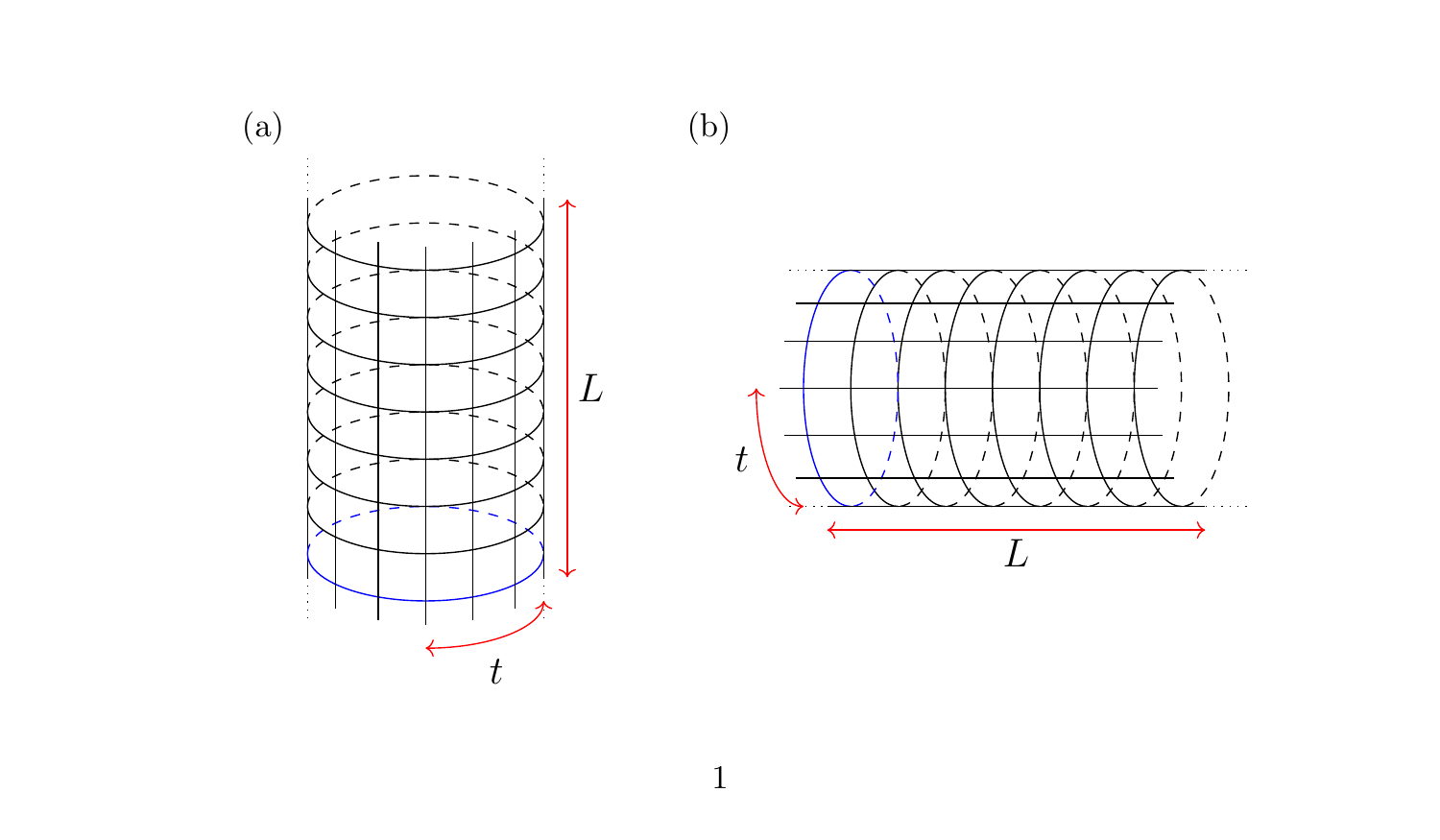}
    \caption{The two ways of computing the partition function of \eqref{part2}. On the right, (b) we take the spatial distance $L=N/D\to\infty$ and the partition function consists of a sum over all states. On the left, (a) we view the system as being on a finite ring $t=2M/D$ but by taking $L\to\infty$ we project onto the maximal eigenvalue only. }
    \label{fig:3}
\end{figure}
Denoting  $\Lambda=\lim_{N\to\infty}\Tr\{\tau'(0)^N \tau'(-c)\}$, the largest eigenvalue, we have,
\begin{eqnarray}\label{Lambda}
\Lambda\!=\prod_{j=1}^{M}\left[\frac{\sinh{(\lambda_j+i\gamma/2)}}{\sinh{(\lambda_j-i\gamma/2)}}\right]^N\frac{\sinh{(\lambda_j-c+i\gamma/2)}}{\sinh{(\lambda_j-c-i\gamma/2)}}~\end{eqnarray}
where the $\lambda_j$ now satisfy,
\begin{eqnarray}\label{TBae}
\left[\frac{\sinh{(\lambda_j-i\gamma/2)}}{\sinh{(\lambda_j+i\gamma/2)}}\right]^{2M}\!=\prod_{j\neq k}^{M}\frac{\sinh{(\lambda_j-\lambda_k-i\gamma)}}{\sinh{(\lambda_j-\lambda_k+i\gamma)}}.
\end{eqnarray}
Note that now any dependence on the impurity parameter $c$ is contained in the eigenvalue, $\Lambda$ rather than the Bethe equations. This has its counterpart, in the language of  conventional TBA \cite{RMP,TWAKM},   where in thermodynamic limit the saddle point of the partition function is found to depend only on the bulk and not the impurity. Taking the log of these Bethe equations \eqref{TBae} we have,
\begin{eqnarray}\label{logBae}
2M\phi(\lambda_j,\gamma/2)&=&-\pi(M-2j+1)+\sum_{k\neq j}^{M}\phi(\lambda_j-\lambda_k,\gamma)
\end{eqnarray}
where the choice of $\mathcal{M}=M$ as well as the logarithmic branches encoded in the successive integers $(M-2j+1), j=1,...,M$ leads to maximal eigenvalue.

It is convenient to rewrite the last equation in terms of  the counting function $Z(\lambda)$,
\begin{eqnarray}\label{count}
Z(\lambda)=2M\phi(\lambda,\gamma/2)-\sum_{k }^{M}\phi(\lambda-\lambda_k,\gamma)
\end{eqnarray}
which has the property that, for $M$ even, $e^{iZ(\lambda_j)}=-1$ when $\lambda_j$ is a solution of \eqref{logBae} and furthermore is an analytic function in the region $|\text{Im}(\lambda)|\leq\text{min}(\gamma/2,\pi-\gamma/2)$. Using these two properties in conjunction with the Residue Theorem we are able to rewrite the sum present in \eqref{count}  as an integral\cite{DDV1,DDV2}
\begin{eqnarray}\nonumber
\sum_{k }^{M}\phi(\lambda-\lambda_k,\gamma)&=&\oint_C\frac{\mathrm{d}\mu}{2\pi i}\phi'(\lambda-\mu,\gamma)\\
&&~~~~~~\times\log{\left(1+e^{-iZ(\mu)}\right)}.
\end{eqnarray}
The contour, $C$ is taken to  run from $-\infty$ to $\infty$ at Im$(\mu)=-\eta$ and the back again at Im$(\mu)=\eta$ with  $0<\eta\leq\text{min}(\gamma/2,\pi-\gamma/2)$. In this way  only  the poles at $\lambda=\lambda_j$ and no other non-analytic points are encircled. Inserting the integral form of the sum into \eqref{count} and rearranging using Fourier transforms and  $2M=tD$ we find a single non linear integral equation (NLIE) which determines $Z$,
\begin{eqnarray}\nonumber
Z(\lambda)=2tD\arctan{\left(e^{\pi \lambda/\gamma}\right)}-2\text{Im}\int_{-\infty}^\infty G(\lambda-\mu-i\eta)\\\label{Zeqn}
\times\log{\left(1+e^{iZ(\mu+i\eta)}\right)}~~~~
\end{eqnarray}
The function $G(x)$ present here is related to the physical two particle phase shift in the AKM and is given by\cite{TWAKM}
\begin{eqnarray}
G(x)=\int \frac{e^{i\omega x}}{4\pi}\frac{\sinh{\left[(\pi/2-\gamma)\omega\right]}}{\cosh{\left[\gamma\omega/2\right]}\sinh{\left[(\pi-\gamma)\omega/2\right]}}.
\end{eqnarray}
 Meanwhile the driving term appearing in the \eqref{Zeqn} is in fact the energy of the fundamental excitation in the AKM. In the universal limit $D\to\infty$ we may replace this with $2tD\arctan{\left(e^{\pi \lambda/\gamma}\right)}\to 2tDe^{\pi\lambda/\gamma}$ ~~\cite{TWAKM}. 
 
 We now return to the eigenvalue $\Lambda$ and seek to express it in terms of the counting function, $Z$.  Taking the logarithm of \eqref{Lambda} we can split $\Lambda$  into a sum of bulk and impurity contributions $\log{\Lambda}=\log{\Lambda_b}+\log{\Lambda_i}$ with  $\log{\Lambda_b}=-iN\sum_j^{M}\phi(\lambda_j,\gamma/2)$ being the bulk part and $\log{\Lambda_i}=-i\sum_j^{M}\phi(\lambda_j-c,\gamma/2)$ being the impurity part. Using the same trick to convert the sum to an integral we have the bulk part is,
\begin{eqnarray}
\log{\Lambda_b}
&=&-itND\int_{-\infty}^\infty \mathrm{d}\lambda \,s(\lambda)\phi(\lambda,\gamma/2)\\\label{bullLambda}&&+2iN\text{Im}\int_{-\infty}^\infty s(\lambda+i\eta)\log{\left(1+e^{iZ(\lambda+i\eta)}\right)}\nonumber
\end{eqnarray}
where we have defined $s(x)=\text{sech}(\pi x/\gamma)/2\gamma$. The impurity part is similarly found to be
\begin{eqnarray}
\log{\Lambda_i}
&=&-it\,D\int_{-\infty}^\infty \mathrm{d}\lambda \,s(\lambda-c)\phi(\lambda,\gamma/2)\\\label{impLambda}&&+2i\text{Im}\int_{-\infty}^\infty s(\lambda-c+i\eta)\log{\left(1+e^{iZ(\lambda+i\eta)}\right)}\nonumber
\end{eqnarray}
At this point we are in a position to perform the Wick rotation $t\to-i\beta$ and obtain the free energy of the AKM for any $\gamma$. We are mostly interested in the region $\gamma\leq\pi/2$ which contains the Toulouse point and the isotropic limit so we will restrict to this which allows us to choose $\eta=\gamma/2$. Making the following definitions, 
\begin{eqnarray}
\epsilon(\lambda)=-iZ\left(\gamma\{\lambda-\log{2\beta D}\}/\pi+i\gamma/2\right)\\
G_0(\lambda)=G(\gamma\lambda/\pi),~~~G_1(\lambda)=G_0(\lambda+i\pi-i0)
\end{eqnarray}
and then taking $D\to\infty$ holding $T_k$ fixed we find from \eqref{impLambda} the impurity part of the free energy is
\begin{eqnarray}\nonumber
F_i(T)=E_{i,0}-\frac{T}{2\pi}\int \text{sech}\left(\lambda+\log{\frac{T}{2T_k}}\right)\left[\log\left(1+e^{-\epsilon(\lambda)}\right)\right.\\\label{impF}
+\left.\log\left(1+e^{-\epsilon^*(\lambda)}\right)\right]~~~~~
\end{eqnarray}
where $E_{i,0}$ is  the dot contribution to the ground state energy and the function $\epsilon(\lambda)$ is a solution of the NLIE,
\begin{eqnarray}\nonumber
\epsilon(\lambda)&=&e^\lambda-G_0*\log\left(1+e^{-\epsilon(\lambda)}\right)\\\label{NLIE}
&&~~~~~~~~~~~~~~~~+G_1*\log\left(1+e^{-\epsilon^*(\lambda)}\right).
\end{eqnarray}
The free energy and the corresponding NLIE have a physically transparent form. The quantity $\text{sech}\left(\lambda+\log{\frac{T}{2T_k}}\right)$ is the density of states of the dot while $\log\left(1+e^{-\epsilon(\lambda)}\right)$  and $\log\left(1+e^{-\epsilon^*(\lambda)}\right)$ can be interpreted as the Fermi functions of the fundamental excitation of the system and its antiparticle where we treat $\epsilon$ and $\epsilon^*$ as the quasi energies. The interactions in the system are encoded in the NLIE which couples the two excitations   together via $G_0(x)$ which is the derivative of the two particle phase shift and $G_1(x)$ which is the derivative of the phase shift between a particle and its anti particle. The driving term $e^\lambda$ is the renormalized excitation energy .

Similarly we can determine the bulk part of the free energy from \eqref{bullLambda}, 
\begin{eqnarray}\nonumber
F_b(T)=E_{b,0}-\frac{T^2L}{2\pi}\int e^\lambda\left[\log\left(1+e^{-\epsilon(\lambda)}\right)\right.\\\label{bulkf}
+\left.\log\left(1+e^{-\epsilon^*(\lambda)}\right)\right].
\end{eqnarray}
Just as for the impurity free energy this takes the form of the density of states $e^\lambda/2\pi$ for the bulk system integrated over the Fermi functions. The similarity between  the expressions for $F_b(T)$ and $F_i(T)$ at low  temperatures is the basis of the Fermi-liquid
description  of the impurity low-temperature physics.

\section{High and low temperature behaviors}
The complete thermodynamics of the AKM in the region $\gamma\leq\pi/2$ are described by \eqref{impF}\eqref{NLIE} and \eqref{bulkf}. We may perform some checks on this result by determining the high and low temperature of the impurity free energy. Before doing this we should examine the expression for the bulk free energy which should describe a gas of non interacting right moving fermions with linear dispersion. Using the lemma proven in \citenum{DDV1} we have that \begin{eqnarray}\label{fbulk}
F_b(T)=E_{b,0}-\frac{\pi T^2}{12}L
\end{eqnarray}
which coincides with the free energy of the non interacting gas computed using standard methods. We may use this result when determining the low temperature behaviour of the impurity free energy. For $\log{(T/2T_k)}\to-\infty$ we can expand the impurity free energy 
\begin{eqnarray}\nonumber
F_i(T)&\approx&E_{i,0}-\frac{T^2}{2\pi T_k}\int e^\lambda\left[\log\left(1+e^{-\epsilon(\lambda)}\right)\right.\\\label{impf}
&&~~~~~+\left.\log\left(1+e^{-\epsilon^*(\lambda)}\right)\right]\\
&=&E_{i,0}-\frac{\pi T^2}{12 T_k}
\end{eqnarray}
which is in agreement with previous calculations of the impurity free energy \cite{TWAKM, RMP}. At high temperature the impurity free energy is determined by the $\lambda\to-\infty$ behaviour of the NLIE. In this limit it is not to difficult to see that $\epsilon=0$ is the solution of \eqref{NLIE} which gives that the high temperature impurity free energy is 
\begin{eqnarray}
F(T\gg T_k)=-T\log{2}
\end{eqnarray}
which describes a decoupled two level system, again in line with known results. 

Such interesting behaviors are also recovered in the Toulouse and isotropic limits. In the Toulouse limit $\gamma=\pi/2$ the phase shifts $G_0(x),~G_1(x)$ vanish and the resonant level model result is recovered. In the other interesting limit $\gamma\to 0$ which gives the isotropic Kondo model the from of the free energy and the NLIE remain unchanged with the exception that we now use the Kondo phase shifts $G_{0,1}(x)\to G^K_{0,1}(x)$ with
\begin{eqnarray}
G_0^K(x)=\int_{-\infty}^\infty\frac{e^{i\omega x}}{4\pi}\frac{e^{-\pi|\omega|}}{1+e^{-\pi|\omega|}}
\end{eqnarray}
and $G_1^K(x)=G^K_0(x+i\pi-i0)$. The high and low temperature expansions derived above remain valid in this case also.

\section{Adding a magnetic field}
It is possible to include a magnetic field, $h$ in the Hamiltonian without breaking integrability. Applying the same procedure as in the previous section we find the following expression for the impurity free energy 
\begin{eqnarray}\nonumber
F_i(T)=E_{i,0}-\frac{T}{4\pi}\sum_{\sigma=\pm}\int \text{sech}{\left(\lambda+\log{\frac{T}{2T_k}}\right)}\\\label{IRLMeps}
\times\left[\log{(1+e^{-\epsilon_\sigma(\lambda)})}+
\log{(1+e^{-\epsilon_\sigma^*(\lambda)})}\right]
\end{eqnarray}
with the newly introduced $\epsilon_\pm(\lambda)$ determined by 
\begin{eqnarray}\nonumber
\epsilon_\pm(\lambda)=e^{\lambda}\mp\frac{\beta h}{1-\gamma/\pi}-G_0*\log{\left(1+e^{-\epsilon_\pm(\lambda)}\right)}\\\label{NLIEIRLMeps}
+G_1*\log{\left(1+e^{-\epsilon_\mp^*(\lambda)}\right)}
\end{eqnarray}
This reduces to the original form when the chemical potential is taken to zero. Here we interpret $\epsilon_+$ and $\epsilon_-$ as the quasi energies of a particle and hole  which couple in opposite fashion to the magnetic field. 

\section{The interacting resonant level model}
In this penultimate section we make some brief remarks on  a model closely related to the AKM, the interacting resonant level model (IRL). This model describes a bath of non interacting chiral spinless fermions which are coupled to localized level on to which they can tunnel. The Hamiltonian is \cite{IRL1}
\begin{eqnarray}\nonumber
H=-i\int \psi^\dag(x)\partial_x\psi(x)+U\psi^\dag(0)\psi(0)d^\dag d\\\label{HIRLM}
+\sqrt{2\Gamma}\left(\psi^\dag(0)d+d^\dag\psi(0)\right)
\end{eqnarray}
where $\psi^\dag(x)$ are right moving spinless fermions, $d^\dag$ describes the localized level, $\sqrt{2\Gamma}$ is the tunnelling strength between the dot and the bulk and $U$ is the strength of the interaction between the occupied level and the bulk.

This model represents a different class of integrable QIMs as in contrast to the AKM there are no internal degrees of freedom and so all the S-matrices in the model are pure phases. Consequently, periodic boundary conditions lead directly to the Bethe equations without the need for a transfer matrix. Despite this the two models are known to be closely related. Indeed using bosonization one can show that the IRLM describes the spin sector of the AKM, but the connection  is less well understood in the Bethe language and so we devote this section to expanding on this. 

The IRLM was solved using coordinate Bethe Ansatz\cite{IRL2} and its thermodynamics were studied using the standard TBA method\cite{CR2}.  We proceed now to present the NLIE approach to the thermodynamics. The eigenstates are  most conveniently written in terms of rapidity variables, $\lambda_j$, related to the  momentum variables $k_j= \mathcal{D}e^{2\lambda_j}$ with the energy of the $N$ particle system  given by $E=\sum_j^N\mathcal{D}e^{2\lambda_j}$. Here $\mathcal{D}$ is the momentum cutoff, having units of energy. The Bethe equations which determine the rapidities are,
\begin{eqnarray}\nonumber
e^{-i\mathcal{D}e^{2\lambda_j}L}\frac{\sinh{(\lambda_j-c'+i\gamma'/2)}}{\sinh{(\lambda'_j-c'-i\gamma'/2)}}
\\\label{Baeirlm}=\prod_{j\neq k}^{N}\frac{\sinh{(\lambda_j-\lambda_k+i\gamma')}}{\sinh{(\lambda_j-\lambda_k-i\gamma')}}
\end{eqnarray}
Herein $\gamma'$ encodes the interaction $U$ and just as in the AKM the relationship is universal only at small interaction $\gamma'\approx \pi/2+U$. The impurity parameter $c'$ is related to the tunnelling $e^{c'}\approx \Gamma/\mathcal{D}$. 

Apart from the exponential on the left hand side these equations are quite similar to those of the AKM which is natural, given the relationship between the two models. We may bring \eqref{Baeirlm} in line with \eqref{BAE} by performing a lattice type regularization of exponential term \cite{Lightcone,longdelayed}
\begin{eqnarray}
e^{-i\mathcal{D}e^{2\lambda_j}L}\rightarrow \left[\frac{\sinh{(\lambda_j-\Theta+i\gamma'/2)}}{\sinh{(\lambda_j-\Theta-i\gamma'/2)}}\right]^{2N-1}
\end{eqnarray}
The equivalence of these two terms is seen by taking $\Theta\to\infty$ such that $\mathcal{D}=(4N-2)\sin{\gamma}e^{-2\Theta}/L$. A similar replacement can be made for the energy and after absorbing $\Theta$ into the Bethe roots we get 
\begin{eqnarray}
E=\frac{1}{L}\sum_j^N\phi(\lambda_j,\gamma'/2)~~~~~~~~~~~~~~~~~~~~~~~~~~~~~~~~~\\\nonumber
\left[\frac{\sinh{(\lambda_j+i\gamma'/2)}}{\sinh{(\lambda_j-i\gamma'/2)}}\right]^{2N-1}\frac{\sinh{(\lambda_j-c'+\Theta+i\gamma'/2)}}{\sinh{(\lambda'_j-c'+\Theta-i\gamma'/2)}}\\
=\prod_{j\neq k}^{N}\frac{\sinh{(\lambda_j-\lambda_k+i\gamma')}}{\sinh{(\lambda_j-\lambda_k-i\gamma')}}~~~~
\end{eqnarray}
which are exactly the equations describing the spin sector of the AKM with $2N-1$ particles.  One may then use all the previous equations for the AKM thermodynamics, \eqref{impF}\eqref{NLIE}\eqref{bulkf} to describe the IRLM with the only difference being that we define the Kondo scale  $T_K=(2N-1)e^{\pi(c'-\Theta)/\gamma'}/L$.

The same type of lattice regularization can be performed on other QIMs without transfer matrices, e.g the Dicke model \cite{Yudson}. The time evolution operator in this regularization is then given by the transfer which produces the same Bethe equations which for the IRL present case coincides with the $\tau(u)$ of the AKM, \eqref{tau}.

\section{Conclusion}
In this paper we have presented a new method for computing the thermodynamic properties of integrable quantum impurity models. The method is based upon use of the transfer matrix of the models and is similar in style to the approaches of previous techniques \cite{DDV1, KP} which were developed for use in quantum field theories and lattice models, but which does not require crossing symmetry which played an important part in previous approaches. We have shown how it can be used to derive the free energy of the anisotropic Kondo model. The result is expressed in terms of a single non linear integral equation which is valid for any $\gamma$. This can be contrasted with the standard approach of the string hypothesis which results in a possibly infinite set of coupled integral equations and is only valid for particular choices of $\gamma$ \cite{takahashi}. 

The method is very general and can be applied to other quantum impurity models such as multi-channel and multi-flavor generalizations of the Kondo model or the Anderson impurity model. It is applicable also to models where the transfer matrix is not apparent as was shown in the case of the interacting resonant level model.

\acknowledgements{CR is supported by the Peter Lindenfeld Fellowship and NA by NSF Grant DMR 1410583}.
\bibliography{mybib}
\end{document}